# Reducing electronic transport dimension to topological hinge states by increasing geometry size of Dirac semimetal Josephson junctions

Cai-Zhen Li,[1,2,#] An-Qi Wang,[2,3,#] Chuan Li,[4,*] Wen-Zhuang Zheng,[2] Alexander Brinkman,[4] Da-Peng Yu,[1] Zhi-Min Liao[2,5,*]

[1]Shenzhen Institute for Quantum Science and Engineering and Department of Physics, Southern University of Science and Technology, Shenzhen 518055, China.

[2]State Key Laboratory for Mesoscopic Physics and Frontiers Science Center for Nano-optoelectronics, School of Physics, Peking University, Beijing 100871, China.

[3]Academy for Advanced Interdisciplinary Studies, Peking University, Beijing 100871, China.

[4]MESA+ Institute for Nanotechnology, University of Twente, 7500 AE Enschede, The Netherlands.

[5]Beijing Key Laboratory of Quantum Devices, and Collaborative Innovation Center of Quantum Matter, Peking University, Beijing 100871, China.

#These authors contributed equally to this work.

* Email: liaozm@pku.edu.cn; chuan.li@utwente.nl

**The notion of topological phases has been extended to higher-order and has been generalized to different dimensions. As a paradigm, $Cd_3As_2$ is predicted to be a higher-order topological semimetal, possessing three-dimensional (3D) bulk Dirac fermions, two-dimensional (2D) Fermi arcs, and one-dimensional (1D) hinge states. These topological states have different characteristic length scales in electronic transport, allowing to distinguish their properties when changing sample size. Here, we report an anomalous dimensional reduction of supercurrent transport by increasing the size of Dirac semimetal $Cd_3As_2$-based Josephson junctions. An evolution of the supercurrent quantum interferences from a standard Fraunhofer pattern to a superconducting quantum interference device (SQUID)-like one is observed when the junction channel length is increased. The SQUID-like interference pattern indicates the supercurrent flowing through the 1D hinges. The identification of 1D hinge states should be valuable for deeper understanding the higher-order topological phase in a 3D Dirac semimetal.**

Topological phases of condensed matter are characterized by the topological invariant-protected bulk-boundary correspondence. This correspondence holds, irrespective of the bulk being gapped or gapless, illustrated by the metallic surface state of a three-dimensional (3D) topological insulator (TI) or Dirac/Weyl semimetal [1-6]. Recently, the topological states of matter have been extended to higher-order topological phases [7-16]. A $D$-dimensional $n$th-order topological phase holds gapless states that with ($D$-$n$)-dimension [17]. The higher-order TIs can be realized in SnTe [11] and bismuth [12]. Similarly, higher-order topological semimetals have been predicted in a 3D Dirac semimetal $Cd_3As_2$ [15,16]. Thus, in $Cd_3As_2$, besides the well-known 3D bulk Dirac states and the 2D Fermi-arc surface states [18-28], there are 1D hinge states on the edges.

Despite the theoretical predication of higher-order hinge states in Dirac semimetal, it is difficult to identify them through transport experiments due to the mixed conduction channels. Here we report the transport modes filter effect in lattice-distorted Dirac semimetal $Cd_3As_2$ nanoplates by constructing Josephson junctions. Each conduction channel, including the bulk Dirac fermions, Fermi-arc surface states, and the topological hinge states, can be distinguished based on their different superconducting coherence lengths by increasing the channel length of the Josephson junction.

For a device with a short junction length, both the bulk and surface state are superconducting coupled, thus the transport in $Cd_3As_2$ contains undistinguishable effects of both bulk and surface state. When increasing the size to exceed the coherent length scales of the bulk state, the supercurrent carried by bulk states will be strongly suppressed [Fig. 1]. Thus, only the surface states remain superconducting coupled because of the longer coherence length benefiting from topological protection and higher mobility [26-28]. Therefore, the electronic transport in the proximity-induced superconducting state is restricted on the surfaces and the effective dimension of the device is reduced from 3D to 2D.

Even further increasing the geometry size, the supercurrent from the large-area 2D

surface can be suppressed. Although the backscattering is forbidden due to the spin-momentum locking property of the topological surface states, there are other finite-angle scatterings in the 2D surface [29], which do not exist in the 1D edge channel. Compared to the 2D surface, the 1D edge channel also undergoes less scattering interaction with bulk states as 1D channel is confined to the edges of the nanoplate. Therefore, the coherence length should be longer in the 1D channel than that in the 2D surface, giving rise to the supercurrent being dominated by the 1D channels located at the intersections between surfaces. The superconducting coupled 1D channels may be the manifestation of 1D hinge states supported by the second-order topological phase. This gives dimensional reduction from 2D to 1D. Such a dimensional reduction of transport carriers can be evidenced by measuring the interference of the supercurrent in a magnetic field. A standard Fraunhofer interference pattern can be observed when a perpendicular magnetic field is applied to a short Josephson junction [Fig. 1(d)]. Upon increasing the junction length, the magnetic field modulation of the critical current start to deviate from the Fraunhofer pattern. By further increasing the junction length to one micron with edge-dominated supercurrent, a SQUID-like pattern can be observed [Fig. 1(e)].

Figure 2(a) shows the scanning electron microscopy (SEM) image of a typical device made from $Cd_3As_2$ nanoplate and Nb electrodes. $Cd_3As_2$ nanoplates were synthesized by chemical vapor deposition, as described elsewhere [30]. Individual $Cd_3As_2$ nanoplates were then mechanically transferred to a highly doped Si substrate with a 285 nm thick $SiO_2$ coating layer, which serves as a back gate. After Ar plasma etching to remove the oxides on the plates, superconducting Nb electrodes were deposited by magnetron sputtering. The junction length (spacing between Nb electrodes) and width are defined as $L$ and $W$, respectively. We have investigated four Josephson junctions with different device size (denoted as devices 1-4, see Supplemental Material, Table S1) [31]. Electrical measurements were performed in a dilution refrigerator with a base temperature of 12 mK.

Below a critical temperature $T_c$ (5K), the junction (device-2 with $L$ = 500 nm)

enters into a superconducting state as shown by the current-voltage (I-V) characteristic in Fig. 2(b). When the *dc* current $I_{dc}$ is small, the source-drain voltage $V_{sd}$ across the junction is zero. While increasing $I_{dc}$ to above a critical current $I_c$, a finite voltage emerges and the junction switches to the normal state. The gate dependence of the normal state resistance $R_N$ [Fig. 2(c)] indicates that the nanoplate is heavily n-doped with the chemical potential located in the conduction band, which is induced by the deposition of Nb electrodes [26]. Figure 2(d) presents the mapping of differential resistance *dV/dI* as a function of $V_g$ and $I_{dc}$. The $I_c$ decreases monotonously when sweeping $V_g$ towards a negative value. As the junction is heavily n-doped (with electron density $\sim 4 \times 10^{18}$ cm$^{-3}$ obtained from the gate dependence of conductivity), the supercurrent is mainly carried by bulk states. Our maximum applied gate voltage is not sufficient to tune the junction across the charge neutrality point. Similar results are also observed in device-1 with $L$ = 300 nm (Supplemental Material, Fig. S1) [31]. An excess current $I_{exc}$ of about 30.5 μA in the I-V trace gives a transparency $D \sim 0.78$ of the junction interfaces (Supplemental Material, Fig. S2) [31].

To study the supercurrent interference, an out-of-plane magnetic field $B$ was applied perpendicular to the substrate. As shown in Fig. 2(e), the field dependence of the critical current $I_c(B)$ shows a regular Fraunhofer pattern with a period of $\Delta B \cong$ 0.34 mT, which can be fitted well using the form $\left| sin\left(\frac{\pi L_{eff} W B}{\Phi_0}\right) / \left(\frac{\pi L_{eff} W B}{\Phi_0}\right) \right|$, where $L_{eff} = L + 2\lambda$, $\lambda$ is the penetration length of the Nb electrodes and is estimated to be about 400 nm from the relation $\Phi_0 = \Delta B (L + 2\lambda) W$. The value of $\lambda$ agrees well with the previously reported values [34]. This interference pattern suggests a nearly uniform supercurrent distribution through the junction, as confirmed by the corresponding supercurrent density profile $J_c(x)$ [inset in Fig. 2(e)] extracted using the Dynes and Fulton approach [35]. See Supplemental Material [30] for details of $J_c(x)$ extraction in Fig. S3. The $I_c(B)$ pattern and corresponding $J_c(x)$ measured in device-1 are provided in Supplemental Material Figs. S4-S5 [31].

As increasing the junction length $L$ to ~ 800 nm (device-3), the gate dependence of $I_c$ as well as the $I_c(B)$ pattern change a lot. The gate dependence of $I_c$ [Fig. 3(a)] exhibits a superconducting dome behavior with a maximum at around $V_g = 0$ V. Such a behavior is different from the monotonous dependence in device-2 [Fig. 2(a)] and consistent with the surface carried supercurrent as previously observed in Cd$_3$As$_2$ nanowire based junctions [26], indicating a dimensional reduction from 3D to 2D. Specifically, for a bulk dominated junction, the induced electrons by positive gate voltage would contribute to the supercurrent and thus leading to a monotonous increase of $I_c$. While for a surface dominated junction, the surface Fermi arcs have a maximum density of states (DOS) at the Dirac point, thus corresponds to a maximum $I_c$ [36,37]. Reduced scatterings from bulk states also facilitate the $I_c$ peak near the Dirac point [26]. Moreover, the surface states dominated superconductivity can also be inferred from the comparison between junction length and coherence length. For a ballistic Josephson junction in which the junction length $L$ is less than the mean free path $l_e$, the coherence length $\xi_0 = \frac{\hbar v_f}{\Delta_i}$ is found to be 460 nm using an approximate Fermi velocity $v_f \simeq 5 \times 10^5$ m/s, and $\Delta_i = 0.72$ meV. While for a diffusive junction ($L > l_e$), the coherence length is $\xi = \sqrt{\xi_0 \frac{l_e}{D}}$, where the transport dimension $D = 3, 2, 1$ for bulk, surface and hinge states, respectively. Using the bulk mean free path $l_e^{bulk} = v_f \frac{m^* \mu_e}{e} = 10$ nm, where $m^* = 0.04 m_e$ and electron mobility $\mu_e = 0.8 \times 10^3$ cm$^2$V$^{-1}$s$^{-1}$ obtained from the measured transfer curve, the corresponding bulk coherence length is $\xi^{bulk} = 40$ nm. While for 2D surface, using an empirical value $l_e^{surface} = 1$ μm obtained from Cd$_3$As$_2$ nanowires [26], the corresponding surface coherence length $\xi^{surface} = 460$ nm. Given its much longer $l_e$ and $\xi$, the surface states tend to have a stronger superconducting coupling with Nb electrodes than the bulk states. Under finite magnetic fields, the weak coupling of bulk states would be further crippled, leading to a superconducting transport that is overwhelmingly dominated by the surface states.

The corresponding $I_c(B)$ pattern of device-3 [Fig. 3(b)] obviously deviates from the standard Fraunhofer pattern [Fig. 3(c)]. There are three main distinct features. First, the half peak width of the central lobe shrinks and is smaller than that of the Fraunhofer fit. Second, the experimental $I_c(B)$ is not strictly periodic, and both the amplitude and periodicity of the interference oscillations deviate from the Fraunhofer fit. Third, considering the parameters of this device, $\Delta B \sim 0.36$ mT from the first minimum of $I_c$, the $L \sim 800$ nm, and the $\lambda \sim 400$ nm, the relation $\Phi_0 = \Delta B(L + 2\lambda)W_{eff}$ gives the effective width of the supercurrent distribution $W_{eff} \sim 3.6$ μm, which is about 0.52 times the junction width $W = 6.9$ μm. Such a reduced effective width of supercurrent distribution suggests a nonuniform supercurrent distribution through the junction, as confirmed by the corresponding $J_c(x)$ [Fig. 3(d)].

By further increasing the junction length to $L \sim 1$ μm (device-4), the $I_c(B)$ shape is greatly changed. As shown in Fig. 4(a), the $I_c(B)$ pattern measured at $V_g = -30$ V (the Fermi level closing to Dirac point in this device) exhibits periodic oscillations with a period $\Delta B \sim 0.18$ mT. This $I_c(B)$ behavior is reminiscent of a SQUID pattern characterized by a simply $\Phi_0$-periodic oscillation, which is usually observed in the configurations containing two Josephson junctions [34,38]. It can also be observed in the systems when the supercurrent is carried by edge states, such as in graphene [39], 2D quantum spin Hall insulator [40-41], and Bi nanowire [12,42-43] based Josephson junctions. In such systems, the two edge channels act as a dc SQUID, but have different origins. The supercurrent edge channel in graphene is attributed to a 'fiber-optic' model in a waveguide cavity as the $\lambda_F$ is large enough [39]. The 2D quantum spin Hall insulators are with insulating bulk and metallic 1D topological edge states that carry the supercurrent [40-41]. The supercurrent edge channel in Bi nanowire is attributed to the higher-order hinge states [12].

Figure 4(b) shows the normalized critical current $I_c(B)/I_c(0)$. The experimental data agrees well with the SQUID-like pattern rather than the Fraunhofer one. The $I_c(B)$ can be fitted well using the SQUID model: $I_c(B) = \sqrt{(I_{c1} - I_{c2})^2 + 4I_{c1}I_{c2}cos^2(\pi\Phi/\Phi_0)}$, where $\Phi = BA_s$ and $A_s \cong (L + 2\lambda)W$ is the effective area enclosed by the two side channels. Two critical currents $I_{c1} = 5.5$ nA and $I_{c2} = 42.5$ nA for the edge channels are obtained from the fitting result. This SQUID-like pattern suggests that the supercurrent is strongly confined to the 1D edges of the junction, as confirmed by the corresponding supercurrent density profile $J_c(x)$ [Fig. 4(c)]. The values of $I_{c1}$ and $I_{c2}$ are within the order of expected critical current for one channel ($\frac{ev_f}{L} \approx$ 80 nA) in the long junction limit. The asymmetry of $I_{c1}$ and $I_{c2}$ should be due to the irregular geometry of the nanoplate or asymmetric coupling strengths between Nb and nanoplate edges. From the Gaussian fit, the width of the edge channel is extracted to be 412 nm and 641 nm for the two edges. The smaller peak emerged right next to left Gaussian peak at around $x$ = -3.4 $\mu$m may indicate an extra edge state at this side of the junction.

The 1D edge channels can be illustrated within the regime of higher-order topological phases. Recently, Călugăru *et al.* [15] and Wieder *et al.* [16] proposed that Cd3As2 is a higher-order Dirac semimetal, in which the higher-order topological phase can support 1D hinge states. Symmetry breaking is believed to facilitate the emergence of higher-order topological phase. Experimentally, for example, structural distortion or uniaxial strain is proposed to turn SnTe into a higher-order TI [11]. Early literatures indicate that the reconstruction of surface atoms has a profound effect on the elastic properties and forms a core-shell structure with different lattice constant between the surface layer and bulk [44-46]. Drawing on this idea, there should be similar surface effects in Cd3As2 nanostructures. In addition, the Cd3As2 nanoplates sitting on a substrate and fixed by electrodes are very susceptible to deformation when cooling down the device to low temperatures due to the mismatch of thermal expansion with the substrate. This deformation, as well as the well-known surface reconstruction effect, may result in the symmetry-breaking and turn Cd3As2 into a higher-order topological semimetal. Besides, the recent study has even pointed out that Cd3As2 intrinsically possesses 1D higher-order hinge (Fermi arc) states as the topological consequence of bulk Dirac nodes [16].

Moreover, similar SQUID-like patterns were observed both near the Dirac point ($V_g$ = -30 V) and far from the Dirac point ($V_g$ = 0 V, see Supplemental Material, Fig. S6) [31]. That is to say, when the junction length is long enough to exceed the coherence length scale of bulk and surface states, the 1D hinge states will always dominate the superconducting transport regardless of the position of Fermi levels. This is due to the longer mean free path of 1D hinge states comparing with 2D surface states, because 1D channel doesn't involve finite-angle scattering process (this could happen in 2D surface states) and undergoes less scattering interaction with bulk states.

It is interesting to note that the oscillation amplitude $\Delta I_c$ (after subtracting the background) is enhanced with increasing magnetic field. As shown in Fig. 4(d), the oscillation amplitude $\Delta I_c$ at 1 mT is three times the value at zero field. Beating in the supercurrent interference between two SQUID channels is responsible for the observed $B$ field enhanced $\Delta I_c$. To obtain the oscillation frequencies of $\Delta I_c$, we perform fast Fourier transformation (FFT) on the $\Delta I_c$ oscillation [Fig. 4(e)]. Two distinct frequencies $F_1 = 5.36$ and $F_2 = 5.78$ 1/mT can be obtained, indicating that two SQUID channels are existed for the observation of the beating mode. If one translates the frequency into corresponding area, it indicates that the two channels located at one side of the junction are separated by 540 nm, comparable to the width of the edge states. Probably, one of the two channels is the intersection of the top and side surface, while the other is the intersection of the bottom and the same side surface (Supplemental Material, Fig. S7) [31].

In conclusion, using superconducting quantum interference, we have demonstrated the transport dimensional reduction from 3D bulk to 2D surface, and then to 1D hinge states by increasing the geometry size in a proximity-induced superconducting $Cd_3As_2$ nanoplate. Our observed supercurrent distributions in one-micrometer-long Josephson junction provide an evidence of 1D hinge states and thus the higher-order topological semimetal in a lattice-deformed $Cd_3As_2$ nanoplate. Our results provide a route to filter out unwanted transport modes, establishing $Cd_3As_2$ as an interesting platform for further studies of Fermi arcs and hinge states.

**Acknowledgements:** This work was supported by National Key Research and Development Program of China (No. 2018YFA0703703 and No. 2016YFA0300802),

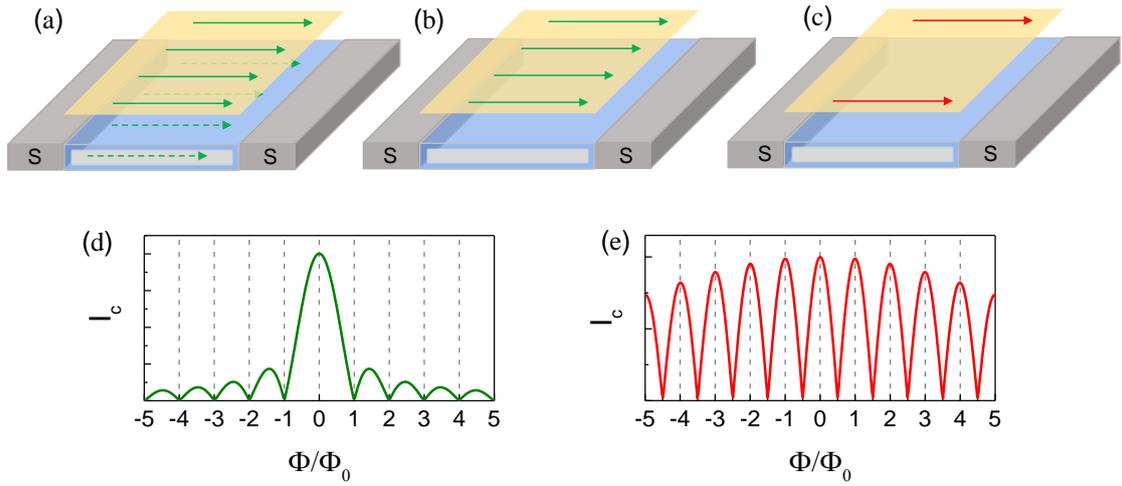

FIG. 1. Supercurrent distribution and interference in a $Cd_3As_2$ based Josephson junction. (a-c) The schematic of a Josephson junction with the supercurrent flows through (a) both bulk and surface, (b) only surface and (c) only the edges of the junction. (d) The Fraunhofer pattern as the supercurrent flows uniformly through the junction. (e) The SQUID-like pattern as the supercurrent flows through the edges of the junction.

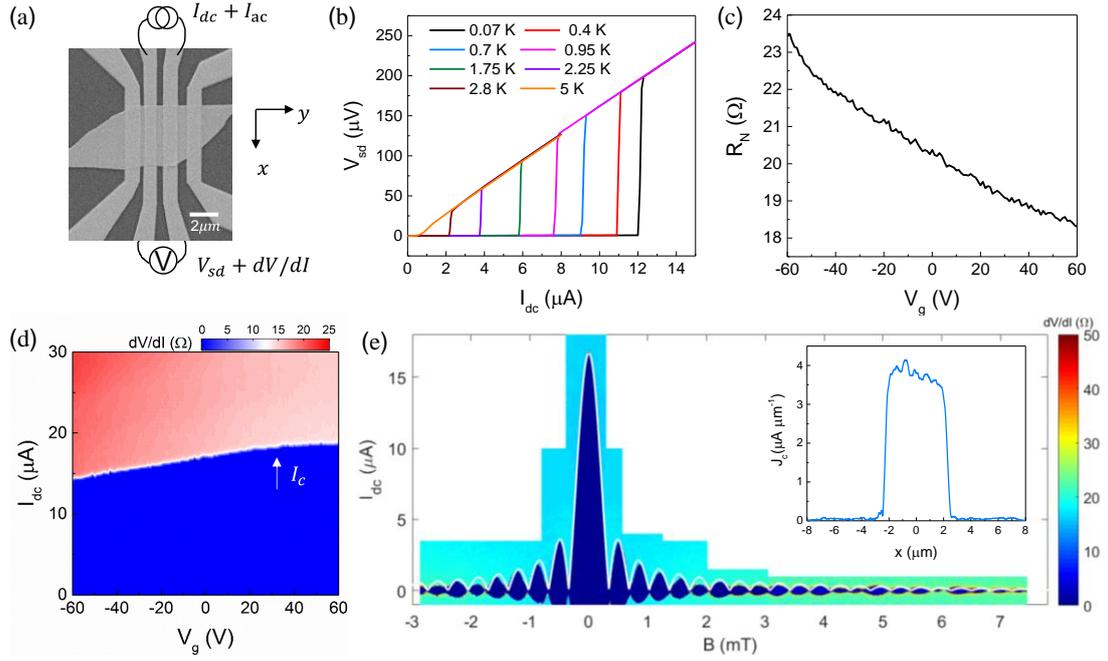

FIG. 2. Characteristics of the Josephson junction device-2 with $L$ = 500 nm, $W$ = 4.62 μm. (a) SEM image of a typical device and the measurement method. $x$ and $y$ denote the spatial coordinates. (b) The I-V characteristics showing the critical current at different temperatures. (c) Normal state resistance $R_N$ vs $V_g$ measured at $I_{dc} = 100$ μA, far above the critical current of the junction. (d) The dV/dI as a function of $V_g$ and $I_{dc}$ at T = 25 mK measured using an ac excitation current $I_{ac} = 100$ nA. The blue area corresponds to the superconducting state, and the white transition boundary corresponds to the critical current $I_c$. (e) The dV/dI as a function of $B$ and $I_{dc}$ at $V_g$ = 0 V. The regular Fraunhofer pattern is observed. A fit with $I_c(B) = I_c(0) \left| \sin\left(\frac{\pi L_{eff} W B}{\Phi_0}\right) / \left(\frac{\pi L_{eff} W B}{\Phi_0}\right) \right|$ is shown as the white curve. Inset: The corresponding supercurrent density profile $J_c(x)$, demonstrating the uniform supercurrent distribution across the junction width.

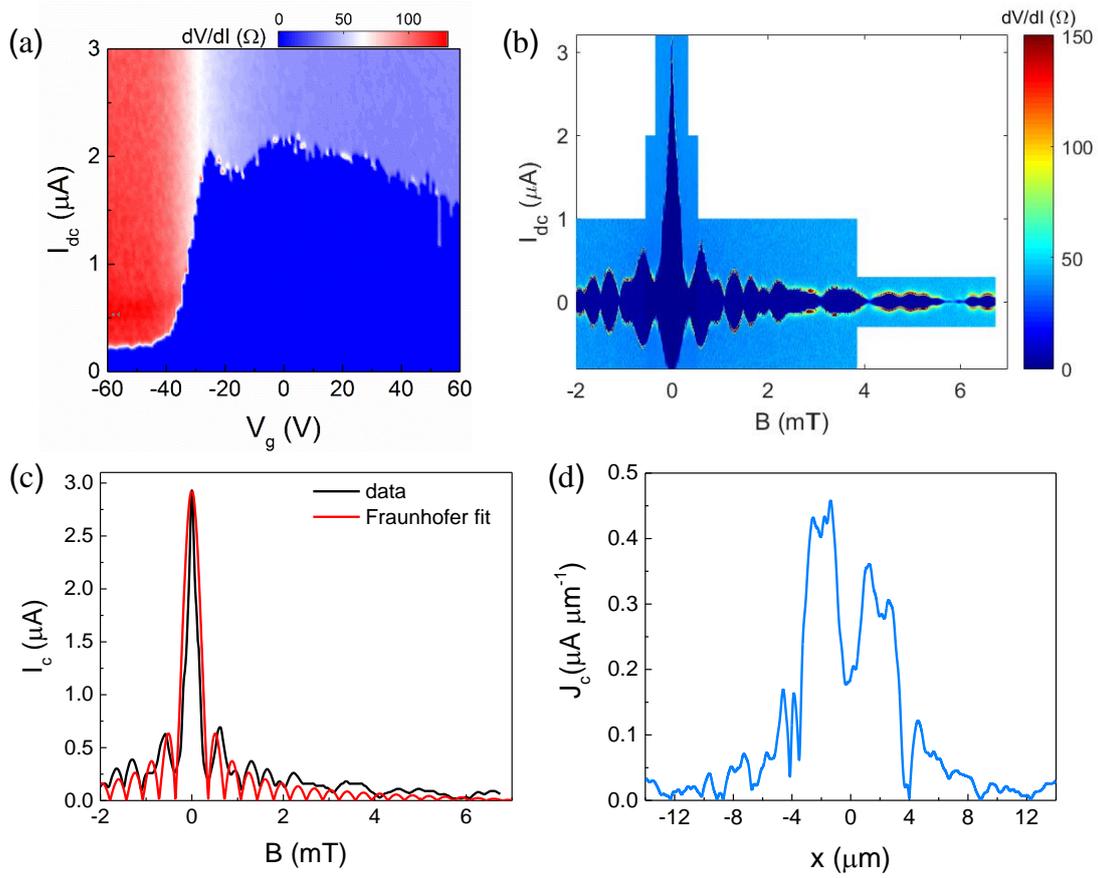

FIG. 3. Characteristics of the Josephson junction device-3 with $L$ = 800 nm, $W$ = 6.9 μm. (a) The *dV/dI* as a function of $V_g$ and $I_{dc}$. (b) The *dV/dI* as a function of $B$ and $I_{dc}$ at $V_g$ = 0 V. (c) The experimental $I_c(B)$ and the Fraunhofer fit. The experimental result apparently deviates from the regular Fraunhofer pattern, indicating that the supercurrent density is not uniform through the junction. (d) The corresponding supercurrent density profile $J_c(x)$ of the experimental $I_c(B)$ data.

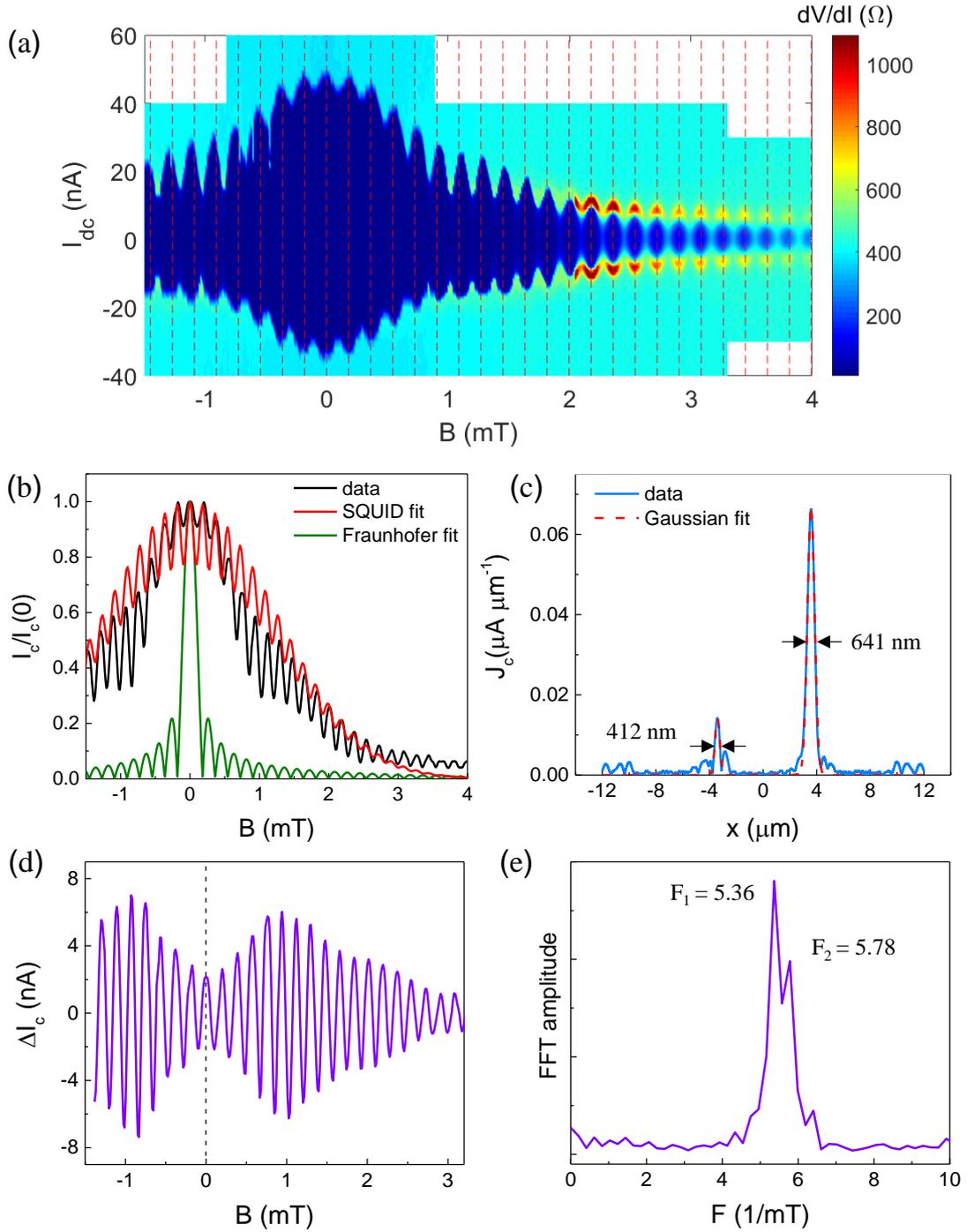

FIG. 4. Characteristics of the Josephson junction device-4 with $L$ = 1 μm. The Fermi level of this device is tuned to near the Dirac point by setting $V_g$ = -30 V. (a) The $dV/dI$ as a function of $B$ and $I_{dc}$. The SQUID-like pattern is observed as marked by the evenly spaced red dotted lines. (b) The comparison of experimental $I_c(B)$ (black curve), Fraunhofer fit (green curve) and the asymmetric SQUID fit (red curve). (c) The corresponding supercurrent density profile $J_c(x)$. (d) The amplitude of the critical current oscillation $\Delta I_c$ after subtracting the background. (e) FFT of the $\Delta I_c$ oscillation with two frequencies $F_1$ = 5.36 and $F_2$ = 5.78 (1/mT).

# Supplemental Material

**Table S1: The device parameters of the measured junctions.**

| device number | device-1 | device-2 | device-3 | device-4 |
|---|---|---|---|---|
| $L$ (nm) | 300 | 500 | 800 | 1000 |
| $W$ (μm) | 4.62 | 4.62 | 6.9 | 7.14 |
| $I_c$ (μA) at 25 mK and $V_g = 0$ V | 25.6 | 16.7 | 3.15 | 0.1 |

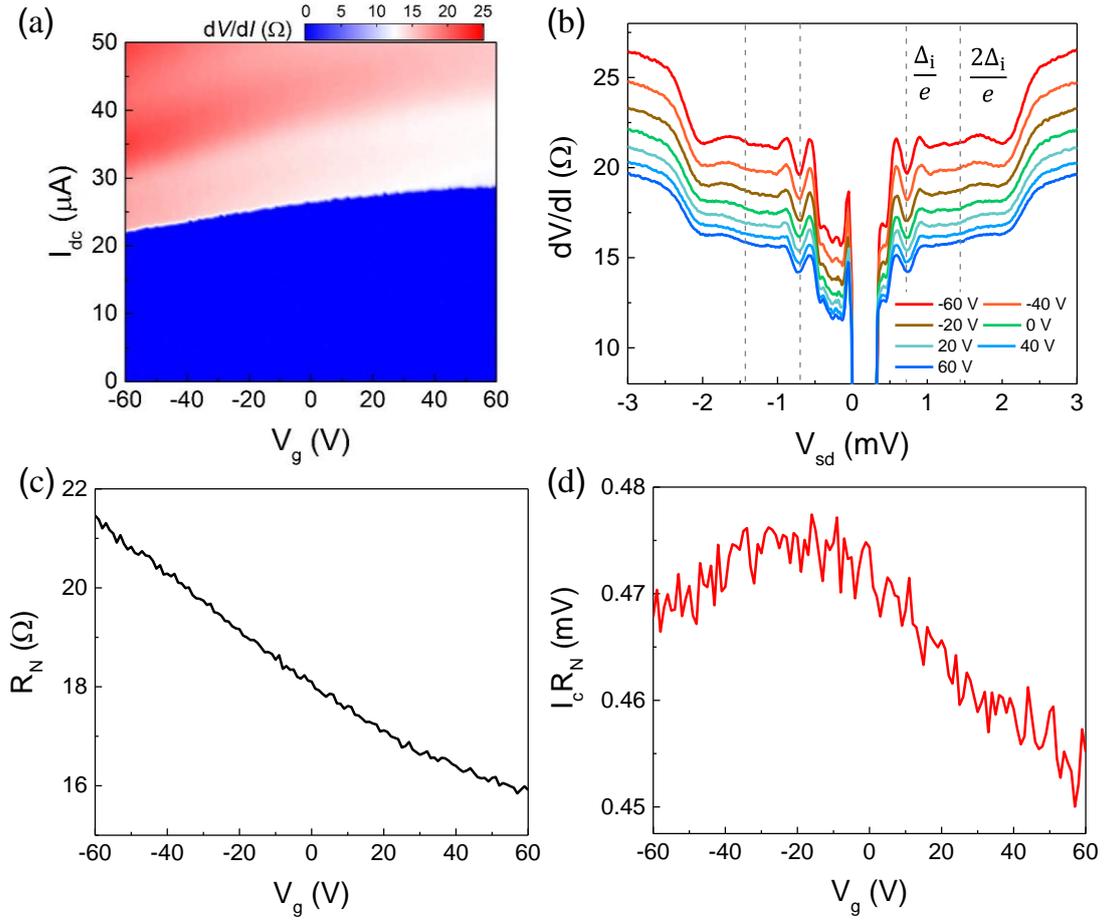

**Figure S1: Characteristics of the Josephson junction device-1 with *L* = 300 nm.**

**a,** The *dV/dI* as a function of $V_g$ and $I_{dc}$, a gate dependent critical current is observd.

**b,** The *dV/dI* as a function of $V_{sd}$ at different gate voltages.

**c,** Normal state resistance $R_N$ as a function of gate voltage $V_g$. $R_N$ is measured at $I_{dc}$ = 100 μA.

**d,** $I_c R_N$ product as a function of gate voltage $V_g$.

We provide in **Fig. S1** the characteristics of device-1. The critical current $I_c$ decreases monotonously when sweeping $V_g$ towards a negative value (**Fig. S1a**), similar to that of device-2. In **Fig. S1b** we show the d*V*/d*I* as a function of the bias voltage ($V_{sd}$) between two superconducting electrodes at different gate voltages. A series of dips in d*V*/d*I* spectra at $V_n = 2\Delta_i/ne$ (n=1,2…) are attributed to the multiple

Andreev reflections. The induced superconducting gap is estimated to be 0.72 meV, which is smaller than the gap value of the Nb layers. The normal state resistance $R_N$ shown in **Fig. S1c** monotonously increases when sweeping $V_g$ towards a negative value, indicating that the nanoplate is heavily n-doped and the chemical potential is located in the conduction band. The n-doping effect is induced by Nb electrodes. The $I_c R_N$ product is shown in **Fig. S1d**. It approachs a max value 0.475 mV at $V_g = -20$ V, yielding a lower bound on the induced gap $I_c R_N \leq \Delta_i/e$. This is compatible with the induced superconducting gap $\Delta_i \sim 0.72$ meV.

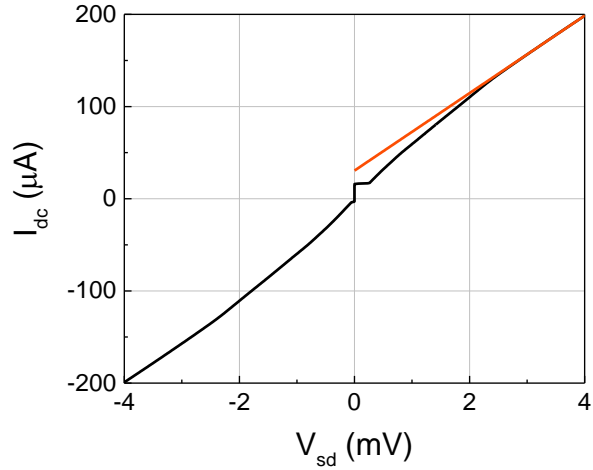

**Figure S2: The I-V trace at large bias measured in device-2 at $V_g = 10$ V. The red curve is the linear fit of I-V trace at high voltage $V_{sd} > 2\Delta_{ind}/e$.**

**Figure S2** shows the I-V trace at large bias. The excess current obtained from the extrapolation of linear fit of I-V trace at high voltage is about $I_{exc} = 30.5$ μA. And the normal state resistance $R_N = 24\ \Omega$ is extracted from the slope of the linear fit. The measured excess current corresponds to $\frac{eI_{exc}R_N}{\Delta_{ind}} = 1.07$ ($\Delta_{ind} = 0.68$ meV in this device), this gives a transparency $D \sim 0.78$ in the Octavio-Tinkham-Blonder-Klapwijk model [31].

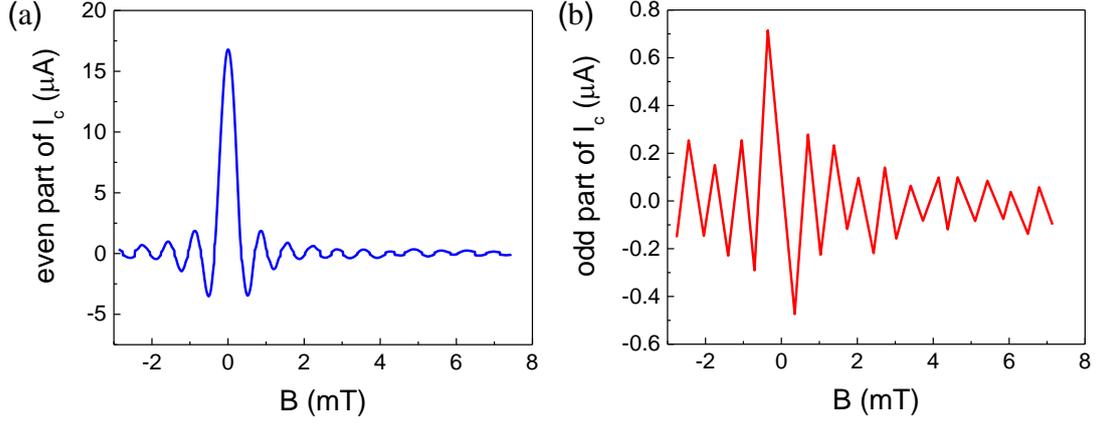

**Figure S3: The extraction of complex critical current $\mathcal{I}_c(B)$.**

**a,** The critical current $I_E(B)$ that corresponds to the even part of the current density profile $J_E(x)$.

**b,** The critical current $I_O(B)$ that corresponds to the odd part of the current density profile $J_O(x)$.

**Method for extraction of supercurrent density profile $J_c(x)$.**

Under a perpendicular magnetic field $B$, the $B$ field modulations of critical current $I_c(B)$ through the junction depends strongly on the supercurrent density profile $J_c(x)$. Specifically, the $I_c(B)$ is given by the complex Fourier transform of $J_c(x)$:

$$I_c(B) = |\mathcal{I}_c(B)| = \left|\int_{-\infty}^{\infty} J_c(x)exp(i2\pi L_{eff}Bx/\Phi_0)dx\right| \quad (1)$$

where $x$ is the dimension along the width of the junction (labeled in **Fig. 2a**), $L_{eff} = L + 2\lambda$ is the effective length of the junction taking the penetration length $\lambda$ into account, and $\Phi_0 = h/2e$ is the flux quantum. Below we extract the $J_c(x)$ from the $I_c(B)$ pattern by employing the Fourier techniques introduced by Dynes and Fulton. In the case of an even current density $J_E(x)$ which represents a symmetric distribution, the odd part of $exp(i2\pi L_{eff}Bx/\Phi_0)$ vanishes from the integral, and equation (1) becomes $\mathcal{I}_c(B) = I_E(B) = \int_{-\infty}^{\infty} J_E(x)cos(2\pi L_{eff}Bx/\Phi_0)dx$. When there is a small but non-vanishing odd component, $J_O(x)$, the odd part of the critical current $I_O(B) = \int_{-\infty}^{\infty} J_O(x) \sin(2\pi L_{eff} Bx/\Phi_0) dx$ should also be considered. Then the complex critical current is:

$$\mathcal{I}_c(B) = I_E(B) + iI_O(B) \tag{2}$$

Therefore, the measured critical current is $I_c(B) = \sqrt{I_E^2(B) + I_O^2(B)}$. The even part $I_E(B)$ is obtained by flipping the sign of every other lobe of the $I_c(B)$ (**Fig. S3a**). The odd part $I_O(B)$ is obtained by interpolating between the minima of $I_c(B)$ and flipping sign between lobes (**Fig. S3b**). The Fourier transform of the resulting complex $\mathcal{I}_c(B)$ yields the supercurrent density distribution:

$$J_c(x) = \frac{1}{\Delta B\, W} \left| \int_{B_{min}}^{B_{max}} \mathcal{I}_c(B) \exp(-i2\pi L_{eff} Bx/\Phi_0) dB \right|. \tag{3}$$

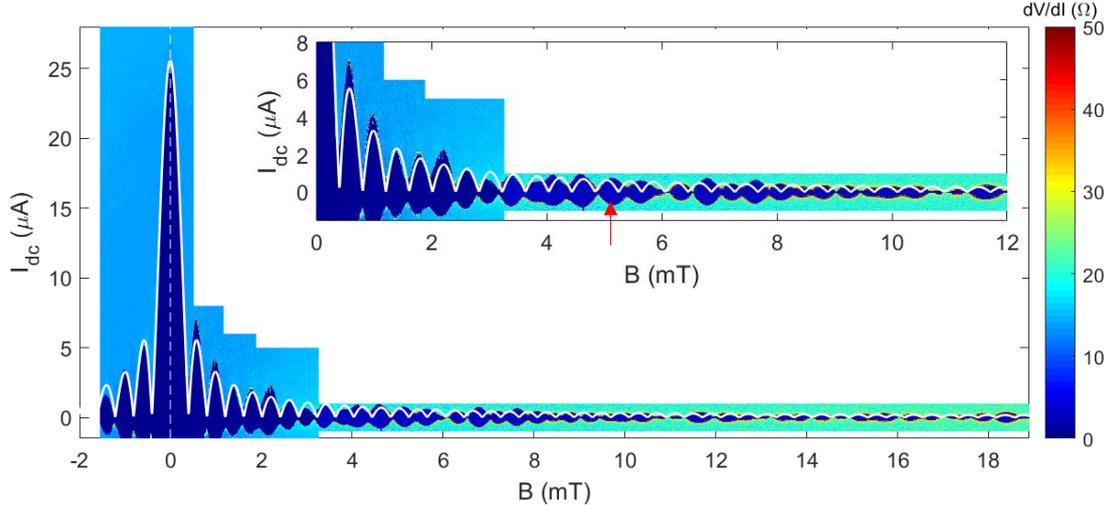

**Figure S4: The $I_c(B)$ pattern of device-1.**

**Main figure,** The *dV/dI* as a function of *B* field and $I_{dc}$ measured in device-1. The mazarine blue regions correspond to the superconducting state and upper transition marks the critical current $I_c$. The white line is the fit to the Fraunhofer pattern formula.

**Inset**, Enlarged *dV/dI* map. The red arrow indicates the field above which the $I_c(B)$ deviates from the standard Fraunhofer pattern.

**Figure S4** shows the $I_c(B)$ pattern measured in device-1. At small *B* fields, a standard Fraunhofer dependence of $I_c(B)$ is observed, indicating a nearly uniformly distributed supercurrent density. While above 5 mT (as the red arrow marked), the $I_c(B)$ deviates from the standard Fraunhofer pattern (inset in **Fig. S4**). Such a deviation can be explained by field-induced orbital decoherence of Andreev pairs. The field required to suppress the Andreev bound states in the bulk is estimated to be as $B^* \sim \Delta_{ind}/eLv_f$ [32]. With the value $\Delta_{ind} = 0.72$ meV (extracted from multiple Andreev reflection) and an averaged $v_f = 0.5 \times 10^6$ m/s, $B^*$ is estimated to be about 4.8 mT, in agreement with the field at which deviation from the Fraunhofer pattern is observed. For $B > B^*$, it is difficult for Andreev reflected electrons and holes to form closed paths to transfer Cooper pairs. In such a case, the bulk carried supercurrent is suppressed [32]. Thus the residual supercurrent is likely to be carried by edge states of the junction. The corresponding supercurrent density distribution $J_c(x)$ (**Fig. S5**) is not as flat as that in device-2 (inset in **Fig. 2e**)), which further indicates the nonuniformity of the supercurrent density.

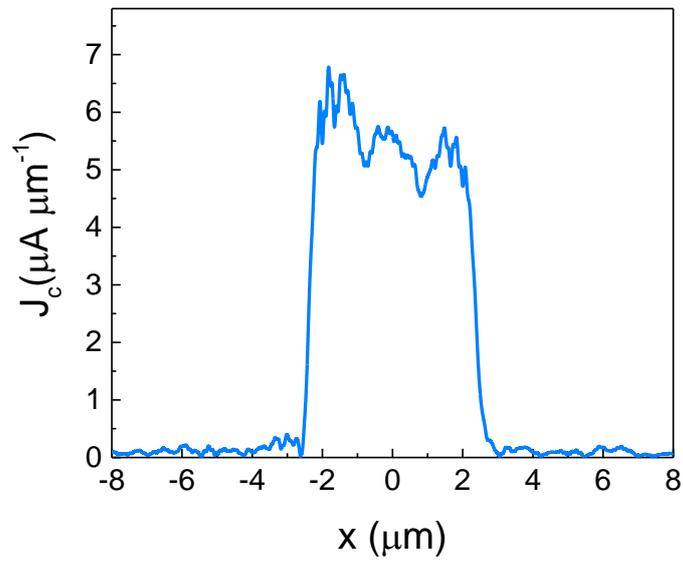

**Figure S5: The supercurrent density profile $J_c(x)$ of device-1.**

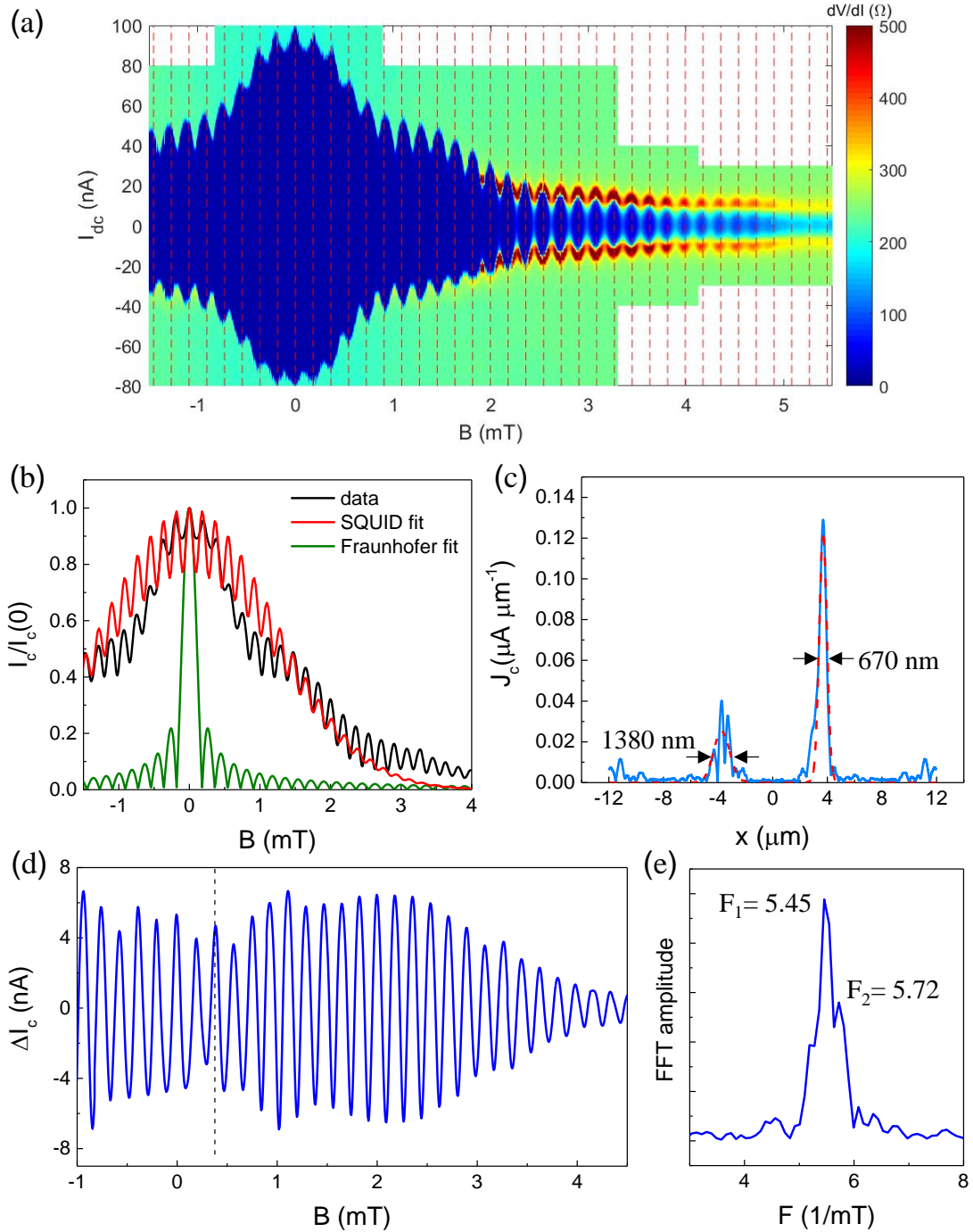

**Figure S6: The $I_c(B)$ pattern of device-4 at $V_g$ = 0 V.**

**a,** The *dV/dI* as a function of *B* and $I_{dc}$. The SQUID-like pattern is observed as marked by the evenly spaced red dotted lines. The oscillation period is the same with that at $V_g$ = -30 V.

**b,** The comparison of experimental $I_c(B)$ displayed as the black curve, Fraunhofer fit displayed as the green curve and the asymmetric SQUID fit displayed as the red curve. For the asymmetric SQUID fit, two different critical currents are obtained: $I_{c1}$ = 11 nA and $I_{c2}$ = 87 nA.

**c,** The corresponding supercurrent density profile $J_c(x)$ of the experimental $I_c(B)$ data. From the Gaussian fit, the width of the edge channel is extracted to be 1380 nm and 670 nm for the two edges.

**d,** The amplitude of the critical current oscillation $\Delta I_c$ after subtracting the background.

**e,** FFT of the $\Delta I_c$ oscillation with two frequencies $F_1$ = 5.45 and $F_2$ = 5.72 (1/mT). If one translates the frequency into corresponding area, it indicates that the two channels located at one side of the junction are separated by 350 nm.

There is a distinct difference between the gate-tuning of interference patterns in Cd3As2 and 2D TIs. In 2D HgTe/HgCdTe and InAs/GaSb quantum wells, when the Fermi energy locates in the bulk gap, the edge-mode superconductivity leads to a SQUID interference pattern, while tuning the Fermi level into the conduction band, there is a Fraunhofer pattern from the bulk superconductivity. In our experiments, similar SQUID-like patterns were observed both near the Dirac point ($V_g$ = -30 V shown in **Fig. 4**) and far from the Dirac point ($V_g$ = 0 V shown in **Fig. S6**), demonstrating the perpetuation of 1D edge channels regardless of the Fermi energy. This indicates a different origination of dimensional reduction from 2D to 1D in Cd3As2 compared to that in 2D TIs. The 1D edge channels in our experiments can be ascribed to the higher-order hinge states.

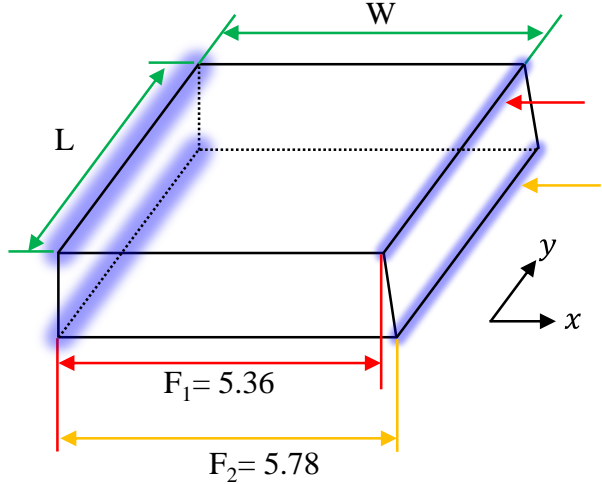

**Figure S7: The schematic diagram showing the superconducting coupled edge states (marked with purple shadows) as well as the two SQUID channels in the junction.** $x$ and $y$ denote the spatial coordinates.

**Figure S7** shows the superconducting coupled edge states and the two SQUID channels in the junction. In the case of irregular geometry at one side of the junction, there is an unaligned separation along the $x$ direction between the two edges, one is the intersection of the top and side surface, while the other the intersection of the bottom and the same side surface (as marked with red and yellow arrow in **Fig. S7**). Such an unaligned separation results in two SQUID channels with different frequencies. If one translates the frequency into corresponding area, it indicates that the two channels located at one side of the junction are separated by 540 nm at $V_g$ = -30 V and 350 nm at $V_g$ = 0 V. It is necessary to note that when the separation is larger than the width of the edge states, these two edges can be distinguished in the supercurrent density profile $J_c(x)$. For example, at $V_g$ = -30 V, the separation 540 nm is larger than the edge width (412 nm), thus an extra smaller peak can be observed at one side of the junction in **Fig. 4c** near $x$ = -3.4 μm. Otherwise, they are difficult to be distinguished as their peaks would superimpose on each other, e g, the peak in **Fig. S6c** near $x$ = -3.7 μm. Nevertheless, they still contribute to two SQUID channels as they are spatially separated.